\documentclass[twocolumn,aps,pre,amsmath,showpacs,tufte-book]{revtex4-1}
\usepackage{graphicx,dcolumn,subfig,caption,bm,amssymb,amsmath,ulem,indentfirst,amsthm}
\usepackage[toc,page]{appendix}
\usepackage{color}

\theoremstyle{plain}
\def\be{\begin{equation}}
\def\ee{\end{equation}}

\newtheorem*{theorem*}{Theorem}

\begin{document}
\author{Bingyu Cui$^{1}$}
\author{Alessio Zaccone$^{1,2,3}$}
\email{az302@cam.ac.uk}

\affiliation{${}^1$Cavendish Laboratory, University of Cambridge, JJ Thomson
Avenue, CB3 0HE Cambridge,
U.K.}
\affiliation{${}^2$Department of Physics ``A. Pontremoli", University of Milan, via Celoria 16, 20133 Milano, Italy}
\affiliation{${}^3$Statistical Physics Group, Department of Chemical
Engineering and Biotechnology, University of Cambridge, Philippa Fawcett Drive,
CB3 0AS Cambridge, U.K.}

\begin{abstract}
A theory of vibrational excitations based on power-law spatial correlations in the elastic constants (or equivalently in the internal stress) is derived, in order to determine the vibrational density of states $D(\omega)$ of disordered solids. The results provide the first prediction of a boson peak in amorphous materials where spatial correlations in the internal stresses (or elastic constants) are of power-law form, as is often the case in experimental systems, leading to logarithmic enhancement of (Rayleigh) phonon attenuation. A logarithmic correction of the form $\sim -\omega^{2}\ln\omega$ is predicted to occur in the plot of the reduced excess DOS for frequencies around the boson peak in 3D. Moreover, the theory provides scaling laws of the density of states in the low-frequency region, including a $\sim\omega^{4}$ regime in 3D, and provides information about how the boson peak intensity depends on the strength of power-law decay of fluctuations in elastic constants or internal stress. Analytical expressions are also derived for the dynamic structure factor for longitudinal excitations, which include a logarithmic correction factor, and numerical calculations are presented supporting the assumptions used in the theory.

\end{abstract}

\title{Vibrational density of states of amorphous solids with long-ranged power-law correlated disorder in elasticity}
\maketitle

\section{Introduction}
Understanding the physics of vibrational spectra of disordered systems is a classical topic in condensed matter physics \cite{Landau1934,Frenkel1946,Zeller1971,Brenig}. Glasses and other disordered solids exhibit anomalous features, compared with their crystalline counterparts. Concerning the thermal properties, at few tens of Kelvin, the specific heat of glasses exhibits an excess over the Debye prediction, in the form of a characteristic maximum in the plot of $C(T)/T^3$. The peak is ascribed to the presence of an excess of states over the Debye density of states (DOS) $\sim\omega^2$, known as the boson peak since its temperature dependence conforms with that of the Bose function, and thus appears to strongly depend on the features of the vibrational modes in the THz frequency \cite{Phillips1981,Frick1995,Angell1995,Greaves2007}. 

Thanks to neutron, X-ray and other inelastic scattering experiments \cite{Sette1998,Benassi1996,Foret2002,Ruocco1999M, Engberg1999,Matic2001,Ruocco2001,Monaco2003,Monaco2006,Chumakov2004,Masciovecchio2006,Monaco2006, Monaco2006E}, computer simulations \cite{Viliani1998,Allen1993,Feldman1993,Feldman1999,Taraskin1997,Taraskin2000,Simdyankin2002, Ribeiro1998,Ribeiro1998,Jund1999,Ruocco2000,Horbach2001}, as well as analytical theory \cite{Buchenau1991,Buchenau1992,Gurevich1993,Gurevich2003,Schirmacher1993,Schirmacher1998, Schirmacher1999, Schirmacher1999P, Schirmacher2000,Horstmann1997,Schirmacher2004,Maurer2004,Mayr2000,Taraskin2002,Kantelhardt2001, Schirmacher2002P, Schirmacher2004P, Maurer2004, Grigera2003, Turlakov2004}, the nature of these excited modes has been widely investigated. Since the boson peak shows up in a frequency range where the broadening of the acoustic excitations becomes of the order of magnitude of resonance frequency, states near the boson peak frequency are neither actually propagating nor localized, and the boson peak itself appears to be closely related to an underlying Ioffe-Regel crossover from ballistic phonon propagation to diffusive excitations, the so-called diffusons~\cite{Tanaka,Allen,BaggioliPRR2019,Baggioli2020}.

Among previous theories, the heterogeneous elasticity theory (HET)~\cite{Schirmacher2004,Schirmacher2006,Schirmacher2007,Cui2019,Lerner2020} uses a field-theoretical scheme to derive the DOS, by assuming  Gaussian uncorrelated spatial fluctuations in the elastic constants of the system \cite{Brenig}. The theory provides a quantitative relation between the boson peak and the Brillouin width (sound attenuation coefficient) $\Gamma$, and reproduces the Rayleigh scattering law $\Gamma \sim \omega^{d+1}$. However, following numerical evidence of a logarithmic enhancement correction of the form $\Gamma(k)\sim-k^{d+1}\ln(k)$ to the Rayleigh scattering law (with wavenumber $k$, in $d$-dimension)~\cite{Gelin2016}, it has been shown analytically that long-ranged power-law spatial correlations in elasticity, or equivalently in the internal stresses, are the cause of such enhancement~\cite{Cui2019}. 

Previous attempts to derive the logarithmic Rayleigh law using HET with power-law correlations in elasticity by Caroli and Lemaitre~\cite{Lemaitre2019} were not successful due to two major simplifying approximations used in their theory, namely the assumption of perfectly isotropic wave propagation (with completely decoupled longitudinal and transverse propagators), which leads to a cancellation of terms and to the vanishing of the logarithmic correction. Caroli and Lemaitre's oversimplifying assumption of isotropic wave propagation is at odds with numerical evidence from Ref.~\cite{Gelin2016}, which showed that wave propagation in the presence of power-law correlated elasticity is locally anisotropic over relatively large length-scales, leading to at least 5 non-vanishing local elastic constants.
In Ref.~\cite{Cui2019}, by finding the rigorous solution to the self-consistent anistropic wave propagation problem, it was possible to derive the logarithmic Rayleigh scattering law, which is ubiquitously observed in experiments and simulations\cite{Kinder1979,Giordano2009,Baldi2010,Baldi2011,Ruta2012,Moriel2019,Flenner,Mizuno2018}, and to show that it is the direct result of the power-law correlation in internal stresses or elastic constants.

Independent evidence supporting the existence of power-law spatially decaying correlations in elasticity have been shown in recent works~\cite{Fuchs2017,Wang2020}. All these facts point towards the importance of properly accounting for long-ranged power-law elastic correlations in the description of the vibrational properties of disordered systems.

A fundamental unanswered question, therefore, is what impact the underlying long-ranged power-law correlations of elasticity (or internal stresses) may have on the DOS. The answer is presented in this article, where we exploit the successful framework of Ref.~\cite{Cui2019} for the acoustic attenuation, and apply it to study the properties of the DOS. We reveal that the boson peak picks up a logarithmic correction which is most evident in the excess DOS. We also show how the boson peak sensitively depends on the strength of power-law correlations of elasticity. We also examine the asymptotic scaling behavior of the DOS in the frequency regimes where modes are quasi-localized due to the disorder (hence undergoing diffusive-like propagation instead of ballistic propagation typical of standard phonons, as demonstrated for glasses in earlier works~\cite{Allen}). 

\section{Theory for longitudinal excitations in 2D}

Numerical simulations, supported by theoretical analysis, and analysis of experimental data, suggest logarithmic enhancement of the Rayleight law in a certain frequency domain~\cite{Gelin2016}. In appendix, we review the model that only predicts the Rayleigh scattering law, such that the mean free path $\ell(\omega)$ scales as $\omega^{-4}$ for small $\omega$.  

According to the theoretical analysis in Ref.~\cite{Cui2019}, such enhancement to Rayleigh scattering of phonons in amorphous solids, originate from long-range power-law spatial correlations of elastic constants or internal stress. For example, the shear stress tensor, $\sigma(\underline{r})=\sigma_0+\Delta {\sigma}(\underline{r})$ is expressed in terms of its mean value plus a random fluctuation, i.e. $\overline{\Delta{\sigma}(\underline{r})}=0$ and $\overline{\Delta{\sigma}(\underline{r}')\Delta {\sigma}(\underline{r}'+\underline{r})}=B(\underline{r})=\kappa^2\cos(4\theta)/(r^2+\zeta^2)$ for some constants $\kappa,\zeta$. The parameter $\kappa$ describes the strength of the disorder, while $\zeta$ controls the regime of frequency where logarithmic enhancement occurs. The resultant 2D self-consistent equations between self-energy $\Sigma(\underline{k},z)$ and the Green's function $G(\underline{k},z)$ for longitudinal waves read \cite{Cui2019}:
\begin{align}
&G(\underline{k},z)=\frac{1}{-z+k^2(c_0^2-\Sigma(\underline{k},z))},\\
&\Sigma(\underline{k},z)=a\int_0^{q_D}\frac{k^2\tilde{B}(\underline{k}-\underline{q})}{-z+q^2(c_0^2-\Sigma(\underline{k},z))}dq^2,\label{eq:self2D}\\
&\tilde{B}(\underline{k})=\int e^{i\underline{k}\cdot\underline{r}}B(\underline{r})d\underline{r},
\end{align}
where the prefactor $a$ is a new parameter reflecting the strength of elastic heterogeneity. Determining $\Sigma(\underline{k},z)$ via solving self-consistent equations above, one can compute the Green's function and hence obtain the density of states (DOS), $D(\omega)$, via the standard Plemelj identity:
\begin{equation}
D(\omega)=\frac{2\omega}{\pi}\text{Im}\{G(z)\},~~ z=\omega^2+i0.
\end{equation}

\begin{figure}[h]
\centering
\includegraphics[width=7.5cm]{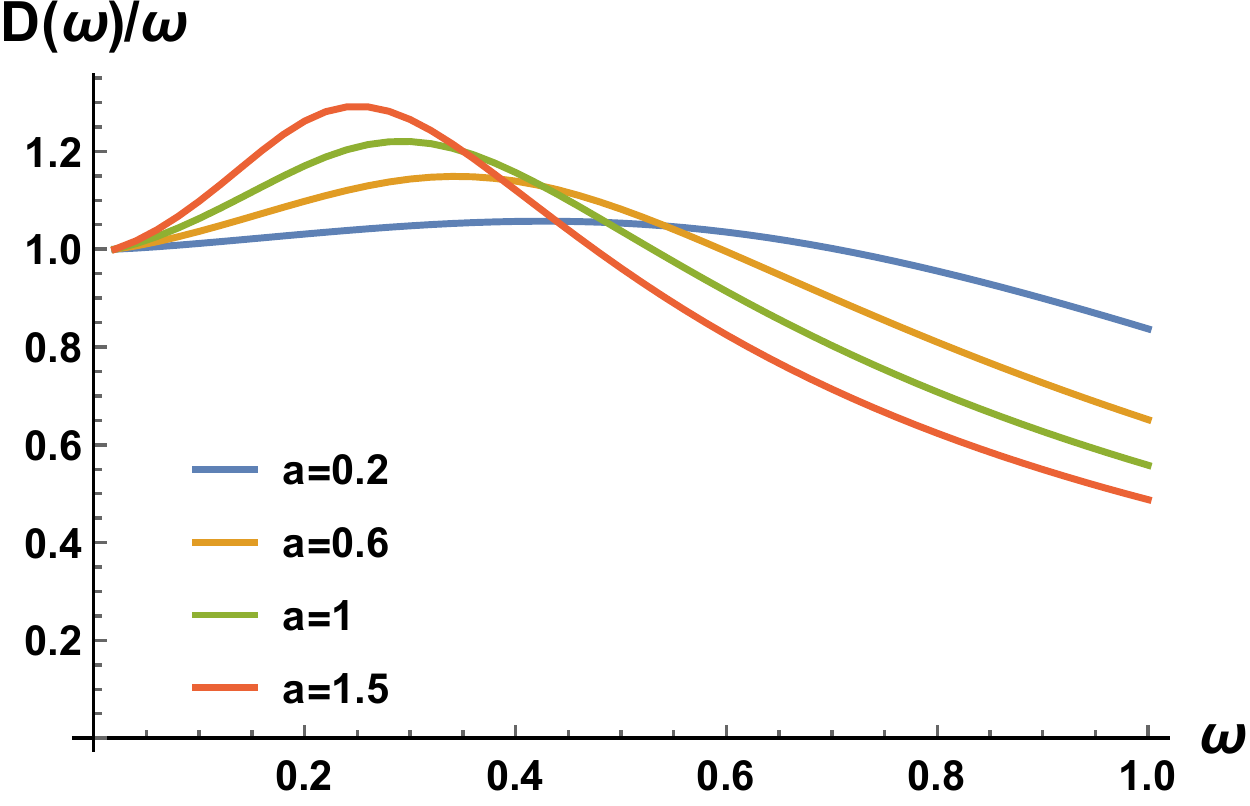}
\caption{Scaled density of states in 2D with fixed parameters $k_D=1,b=0.1, c_0=0.5$. The height of curve has been rescaled.}
\label{fig:selfequa}
\end{figure}

\begin{figure}[h]
\centering
\includegraphics[width=7.5cm]{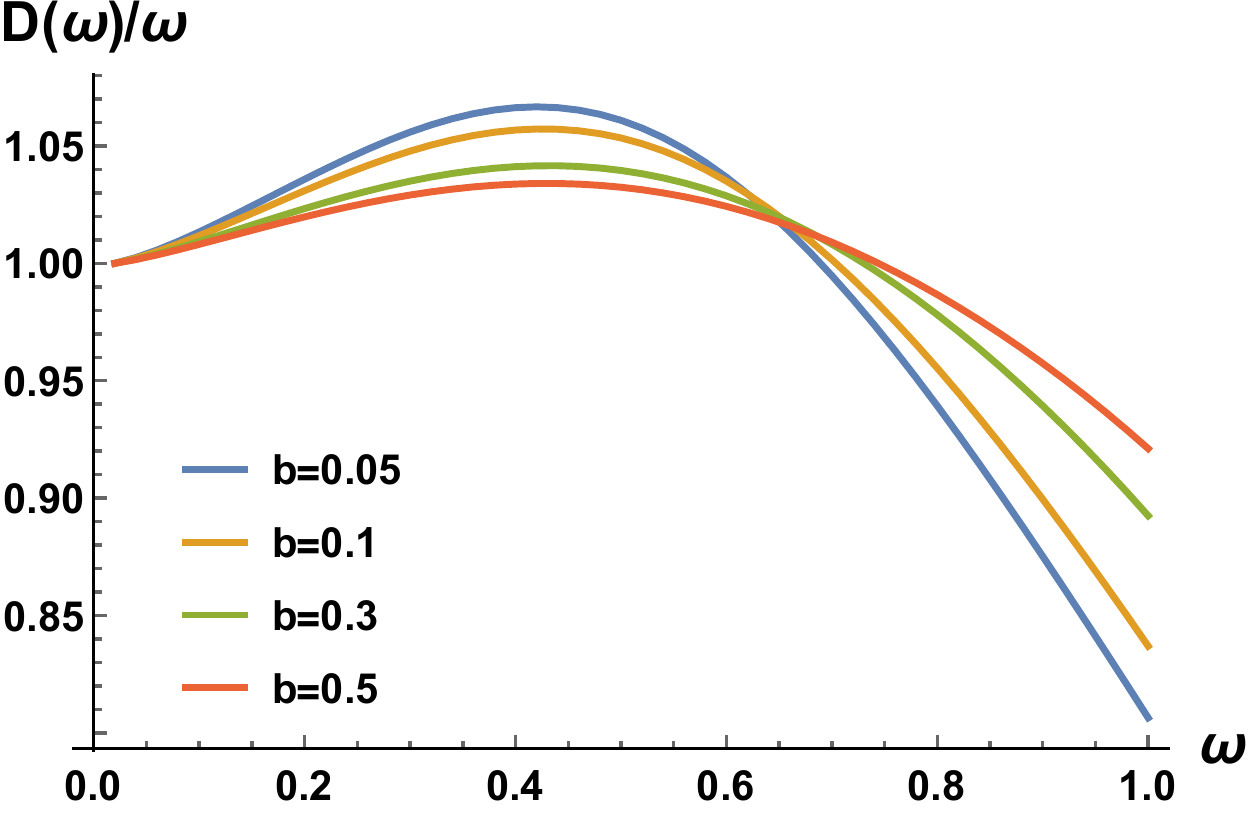}
\caption{Scaled density of states in 2D with fixed parameters $k_D=1, a=0.2, c_0=0.5$. The height of curve has been rescaled.}
\label{fig:selfequb}
\end{figure}

From Ref. \cite{Cui2019}, we can approximate $\tilde{B}(\underline{k}-\underline{q})\sim-\ln(bk)$, which is valid upon assuming the linear (acoustic) dispersion relation between wavenumber $q$ and frequency $\omega$. The parameter $b$ depends on $\zeta$ in $B(\underline{r})$, in a way such that the larger $\zeta$ is, the larger $b$ turns into, thus the lower frequency regime that log-effect emerges. Substituting this into Eq. \eqref{eq:self2D} and re-introducing the parameter $a$, we solve the self-consistent equation of self-energy $\Sigma(z)$ in 2D:
\begin{equation}
\Sigma(z)=a\int_0^{q_D}\frac{-zq\ln(zb)}{-z+q^2(c_0^2-\Sigma(z))}dq,
\end{equation}
from which the scaled DOS is obtained as
\begin{equation}
\frac{D(\omega)}{\omega}\propto-\text{Im}\int_0^{k_D}\frac{kdk}{-\omega^2+k^2(c_0^2-\Sigma(\omega^2))}.
\end{equation}
Figure \ref{fig:selfequa} and Figure \ref{fig:selfequb} are plots of DOS with different $a$ and $b$, scaled by the 2D Debye law $\sim\omega$. From Fig. \ref{fig:selfequa}, we find that the boson peak becomes flatter with smaller $a$ values, which indicates weaker disorder in elasticity. Looking at Fig. \ref{fig:selfequb}, the boson peak shifts to lower frequency and becomes stronger when $b$ decreases. Also, it is clear that the peak due to the contribution of longitudinal prorogation mode is flat, a result demonstrated in previous work \cite{Baggioli2019}. 

\section{General theory for amorphous solids in 3D}
The model in the last section describes purely longitudinal waves in 2D and is served to illustrate the basic functioning of the theoretical framework on an easy example. Now we present the full theory, with the inclusion of transverse waves, in 3D. Making similar assumptions of elastic disorder as in Refs. \cite{Schirmacher2006,Schirmacher2007}, we consider an elastic medium with a mass density $m_0$, shear modulus $G$, bulk modulus $K=\lambda+2G/3$ where $\lambda$ is the longitudinal Lam\'{e}'s constant. The elastic constants are related to the longitudinal and transverse local sound velocities as $c_T\equiv G/m_0$, $c_L^2\equiv (K+4G/3)/m_0=(\lambda+2G)/m_0$, respectively. The Lam\'{e}'s constant is set to be $\lambda=\lambda_0$, while the shear modulus includes a random spatial variation $G(\underline{r})=G_0[1+\Delta{G}(\underline{r})]$. The random function $\Delta{G}(\underline{r})$ is supposed to have a long-ranged power-law decay $\overline{\Delta{G}(\underline{r}')\Delta{G}(\underline{r}'+\underline{r})}=B'(\underline{r})\propto\gamma^2/(r^2+\xi^2)^{3/2}$ for some constants $\gamma$ and $\xi$, in agreement with recent evidence for glasses and granular materials~\cite{Fuchs2017,Wang2020}. The explicit form of angular component in $B(\underline{r})$ is not relevant to results and is not shown in the last expression. The self-consistent Born approximation for the complex self-energy $\Sigma(\omega)$, based on the standard replica-trick, leads to the set of self-consistent equations:
\begin{align}
&\Sigma(\omega)=g\int_0^{q_D}-\omega^2q^2\ln(\omega b)[G_L(q,\omega)+G_T(q,\omega)]dq,\notag\\
&G_L(k,\omega)=\frac{1}{-\omega^2+k^2(c_L-2\Sigma(k,\omega))},\\
&G_T(k,\omega)=\frac{1}{-\omega^2+k^2(c_T-\Sigma(k,\omega))}.\notag
\end{align}
We again assume the linear dispersion relation between $k$ and $\omega$, which is verified by dynamical structure factor calculations in Appendix A. Likewise, the new parameter $g$, which absorbs the disorder-strength parameter $\gamma$, is the prefactor of the self-consistent equation for 
$\Sigma(\underline{k},\omega)$. The Debye length and frequency are given by $k_D^{-1}$ and $\omega_D=c_Dk_D$, with $c_D=[1/3\left((c_L + \text{Re}[\Sigma(0)])^{-3} + 2(c_T + \text{Re}[\Sigma(0)])^{-3}\right)]^{-1/3}$ \cite{Schirmacher2007}.
The DOS can be calculated as
\begin{equation}
D(\omega)\propto\omega\int_0^{k_D}k^2[G_L(k,\omega)+2G_T(k,\omega)]dk.
\end{equation}

\begin{figure}[t]
\subfloat[][]{
\begin{minipage}[t]{0.6\textwidth}
\flushleft
\includegraphics[width=0.8\textwidth]{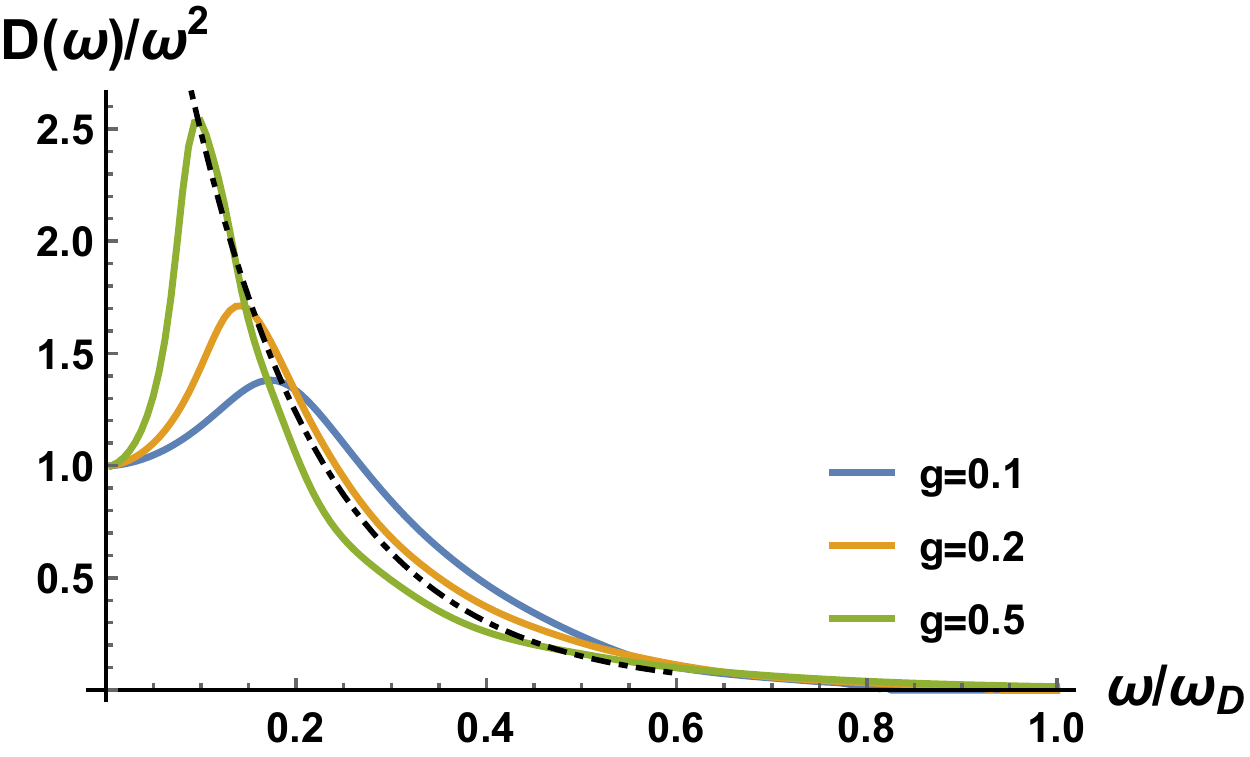}
\end{minipage}
}
\hfill
\subfloat[][]{
\begin{minipage}[t]{0.6\textwidth}
\flushleft
\includegraphics[width=0.8\textwidth]{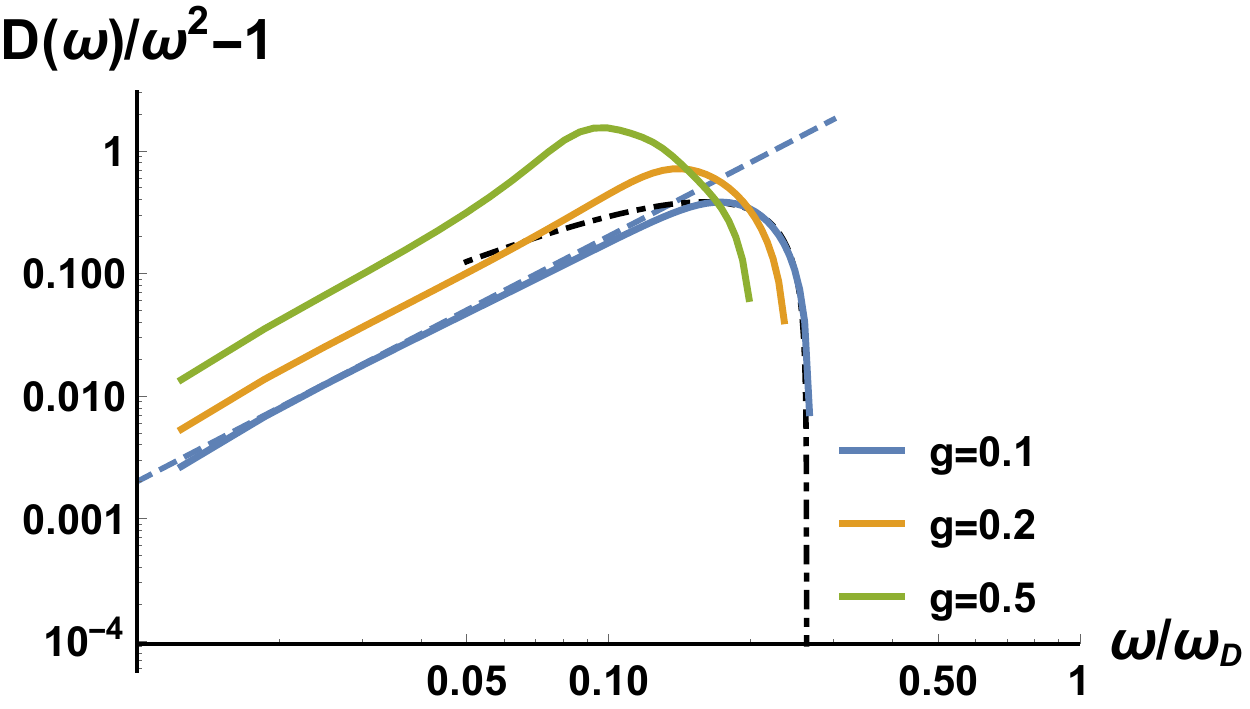}
\end{minipage}
}
\caption{Plot of DOS with fixed parameters $k_D, c_T, c_L,b$ being $1,0.5,1,0.1$ respectively. The height of curve has been rescaled. Panel (a): reduced DOS, $D(\omega)/\omega^2$, for different $g$ values; the dot-dashed line is a simple exponential trend line. Panel (b): reduced excess DOS, $D(\omega)/\omega^2-1$, for different $g$ values. The dashed line indicates a $\sim\omega^{4}$ scaling in the DOS. The dashed-dotted line indicates a logarithmic $-\omega^{2}\ln\omega$ trend about the boson peak frequency.}
\label{fig:self3Dg}
\end{figure}

In Fig. \ref{fig:self3Dg}, we show the typical reduced DOS $D(\omega)/\omega^2$, i.e. the usual boson peak representation, as well as the reduced excess DOS, $D(\omega)/\omega^2-1$, against the $\omega/\omega_D$. It can be seen from Fig. \ref{fig:self3Dg} (a) that the boson peak frequency decreases sharply upon increasing $g$, hence upon increasing the degree of elastic disorder, $\gamma$. Hence, larger disorder lifts up the boson peak and shifts it to lower frequencies, in accordance with earlier findings from simulations~\cite{Milkus2016}. We also note from Fig. \ref{fig:self3Dg}(a) that the boson peak drops exponentially with its frequency $\omega_{BP}$ upon increasing $g$, which is a new law found here by our theory. This might be related to the exponential decaying mode in spectra of activation energies in metallic glasses~\cite{Wei2019}. In Fig. \ref{fig:self3Dg}(b), obviously, the excess over the Debye level is different form zero only above a certain frequency threshold. The excess DOS turns out to vanish as $\omega^4$ for $\omega\rightarrow0$. Upon approaching the boson peak frequency, i.e. $\omega\lesssim\omega_{BP}$, where the DOS $D(\omega)$ displays the $\omega^2$ dependence, the excess DOS tends to flatten out. The nature of vibrational eigenmodes varies as $\omega$ changes. In particular, the asymptotic behaviour $\sim\omega^4$ as $\omega\rightarrow0$, is consistent with results of previous work using HET and Gaussian disorder in elastic constants in \cite{Schirmacher2006}, and with numerical evidence in \cite{Mizuno2018,Silbert2005, Silbert2009, Charbonneau2016, Mizuno2017}. Remarkably, when the frequency becomes comparable to the boson peak frequency, an additional trend representing a logarithmic correction dependency is observed, as shown in Fig. 3(b). This is the first prediction of the logarithmic correction in the reduced DOS, which is expected based on the generic direct proportionality relation between the excess DOS and the phonon attenuation coefficient highlighted in Refs. \cite{Schirmacher2007, Mizuno2018}.

\begin{figure}
\centering
\subfloat[][]{
\begin{minipage}[b]{0.5\textwidth}
\includegraphics[width=0.8\textwidth]{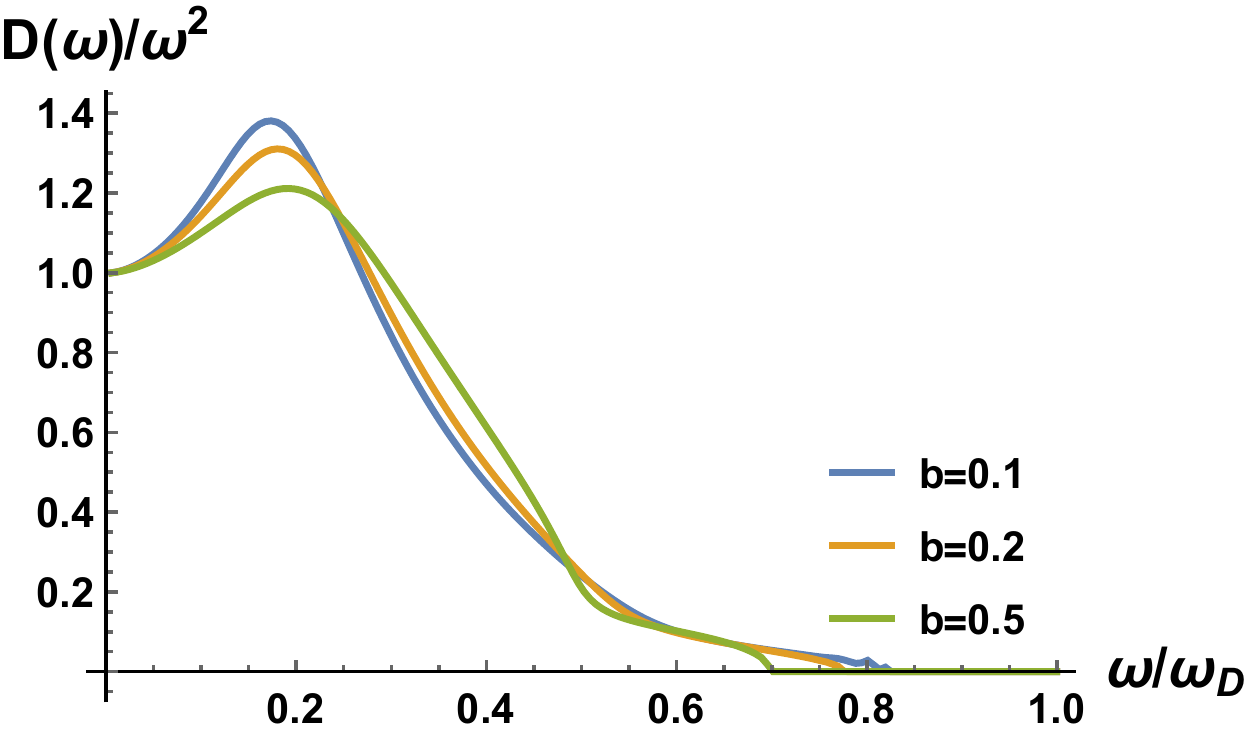}
\end{minipage}
}\\
\subfloat[][]{
\begin{minipage}[b]{0.5\textwidth}
\includegraphics[width=0.8\textwidth]{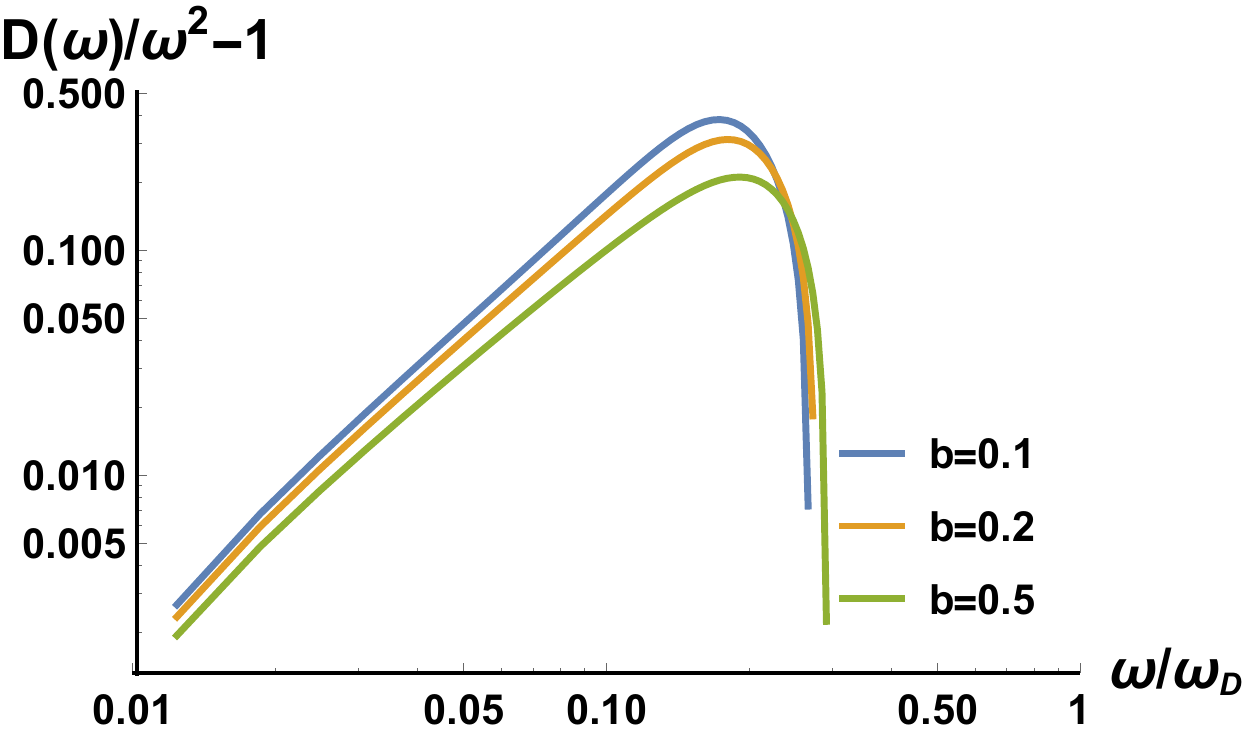}
\end{minipage}
}
\caption{Plot of reduced density of states (a) and reduced excess DOS (b) with fixed parameters $k_D (1.0), c_T(0.5), c_L(0.1),g(0.1)$ and varying $b$. The height of curve has been rescaled. Panel (a): Reduced DOS, $D(\omega)/\omega^2$, for different $b$. Panel (b): Reduced excess DOS, $D(\omega)/\omega^2-1$, for different $b$. The dashed line indicates a $\sim\omega^{4}$ scaling in the DOS.}
\label{fig:self3Db}
\end{figure}

We also show how the boson peak changes with different values of $b$ in Fig. \ref{fig:self3Db}(a), where a similar monotonic relation as in the purely longitudinal case is observed. In Fig. \ref{fig:self3Db}(b), clearly the excess $D(\omega)-\omega^2$ is reduced upon increasing $b$ because of the interplay between the prominent $\sim\omega^4$ behavior at lower $\omega$ and the influence of logarithmic enhancement at higher $\omega$.

\section{Conclusions}
In summary, we developed a theory of vibrational excitations in disordered media with long-ranged power-law correlated disorder to extract the density of states (DOS). The assumption of power-law correlated disorder 
in elastic properties (elastic constants or stresses), supported by evidence found in glasses~\cite{Fuchs2017} and granular materials~\cite{Wang2020}, has been key to derive the logarithmic enhancement $\sim -k^{d+1}\ln k$ of Rayleigh scattering in glasses in our previous work~\cite{Cui2019}, by accounting for the anisotropic character of wave propagation in the solid locally.
The theory reproduces the boson peak in the DOS along with its dependence on the strength $\gamma$ (or $g$), and on the characteristic scale $b$, of power-law correlated disorder. Importantly, the theory predicts 
a logarithmic correction $\sim -\omega^{2}\ln \omega$ visible in the reduced excess DOS around the boson peak frequency, predicted here for the first time.
The theory also predicts that the boson peak decays exponentially with its frequency $\omega_{BP}$ upon increasing the strength of disorder $g$, which might be related to evidence recently found in metallic glasses~\cite{Wei2019}.
The theory predicts the existence of a $\sim\omega^4$ regime (in $3D$) at low frequency below the boson peak, which can be ascribed to Rayleigh scattering. Similar $\omega^{4}$ modes have been recently discovered in the non-Debye part of the spectrum, which may be ascribed to localized anharmonic modes~\cite{Lerner1,Lerner2}, which have been demonstrated to be universal in recent work~\cite{Lerner3,Zapperi}. It appears that the present theory, which is rather on the continuum level and does not account for anharmonicity, cannot predict those modes, while the predicted $\omega^{4}$ refers most likely to Rayleigh scattering since we have checked that for a $2D$ system the scaling is much closer to $\omega^{3}$.
The logarithmic feature in the reduced DOS predicted by this theory calls for more detailed investigation of experimental data in future analysis.
Studying the influences of anharmonicity~\cite{Baggioli2019,Mossa}, nonaffine elasticity~\cite{Zaccone2013,Laurati,Cui2020} as well as glass stability~\cite{Flenner} on the vibrational excitation modes within the current theoretical model, will be the object of future work.

\begin{acknowledgements}
Useful discussions with E. Lerner and E. M. Terentjev are gratefully acknowledged. This work was supported by the CSC-Cambridge Scholarship (B.C.) and by the US Army ARO Cooperative Agreement W911NF-19-2-0055 (A.Z.).
\end{acknowledgements}

\begin{appendix}
\section{Longitudinal waves with Gaussian disorder in elastic constant}
The self-consistent Born approximation, using the replica trick to evaluate the Green's function of an elastic Lagrangian with quenched Gaussian disorder in the elastic constant, was proposed in Ref. \cite{Schirmacher2004}. This leads to a self-consistent relation between the (complex) self-energy $\Sigma(z)$ and the 3D Green's function $G(z)$ of the longitudinal waves as:
\begin{align}
&\Sigma(z)=\frac{\gamma}{2}\sum_{|\underline{k}|<k_D}\frac{k^2}{-z+k^2(c_0^2-\Sigma(z))},\notag\\
&G(z)=\sum_{|\underline{k}|<k_D}\frac{1}{-z+k^2(c_0^2-\Sigma(z))},
\end{align}
where parameters $c_0,\gamma$ correspond to mean of the sound velocity and to the variance of the elastic autocorrelations, respectively. The Debye wavenumber is $k_D=(6\pi^2N/V)^{1/3}$ for a system with $N$ particles and volume $V$, so that the frequency $z=\omega^2+i0$ has a (Debye) cut-off value at $\omega_D=c_0k_D$. With standard identification $1/N\sum_{|k|<k_D}\rightarrow(3/k_D^3)\int_0^{k_D}k^2dk$, we can transform the discrete sums into continuous integrals over momentum space:
\begin{align}
&\Sigma(z)=\frac{3N\gamma}{2k_D^3}\int_0^{k_D}\frac{k^4}{-z+k^2(c_0^2-\Sigma(z))}dk,\notag\\
&G(z)=\left(\frac{3N}{k_D^3}\right)\int_0^{k_D}\frac{k^2}{-z+k^2(c_0^2-\Sigma(z))}dk.
\label{eq:appendselfcon}
\end{align}
The integral in Eq. \eqref{eq:appendselfcon} can be calculated analytically, giving
\begin{align}
&\int_0^{k_D}\frac{k^4}{-z+k^2(c_0^2-\Sigma(z))}dk\notag\\
=&\frac{1}{(c_0^2-\Sigma)^2}\int_0^{k_D}\frac{[(c_0^2-\Sigma)k^2-z](c_0^2-\Sigma)k^2+(c_0^2-\Sigma)k^2z^2}{(c_0^2-\Sigma)k^2-z}\notag\\
=&\frac{1}{(c_0^2-\Sigma)^2}\int_0^{k_D}(c_0^2-\Sigma)k^2+\frac{(c_0^2-\Sigma)k^2z-z^2+z^2}{(c_0^2-\Sigma)k^2-z}dk\notag\\
=&\frac{1}{(c_0^2-\Sigma)^2}\int_0^{k_D}\left[(c_0^2-\Sigma)k^2+z+\frac{z^2}{(c_0^2-\Sigma)k^2-z}\right]dk\notag\\
=&\frac{k_D^3}{3(c_0^2-\Sigma)}+\frac{zk_D}{(c_0^2-\Sigma)^2}+\frac{z^2}{(c_0^2-\Sigma)^{5/2}}\ln\left|\frac{(c_0^2-\Sigma)^{1/2}k-\sqrt{z}}{(c_0^2-\Sigma)^{1/2}k+\sqrt{z}}\right|.
\end{align}
Setting $k_D,c_0,\gamma$ and a proper initial value of self-energy, $\Sigma_0(z)$, we can use an iteration scheme to numerically determine $\Sigma(z)$. 

\begin{figure}[t]
\centering
\includegraphics[width=7.5cm]{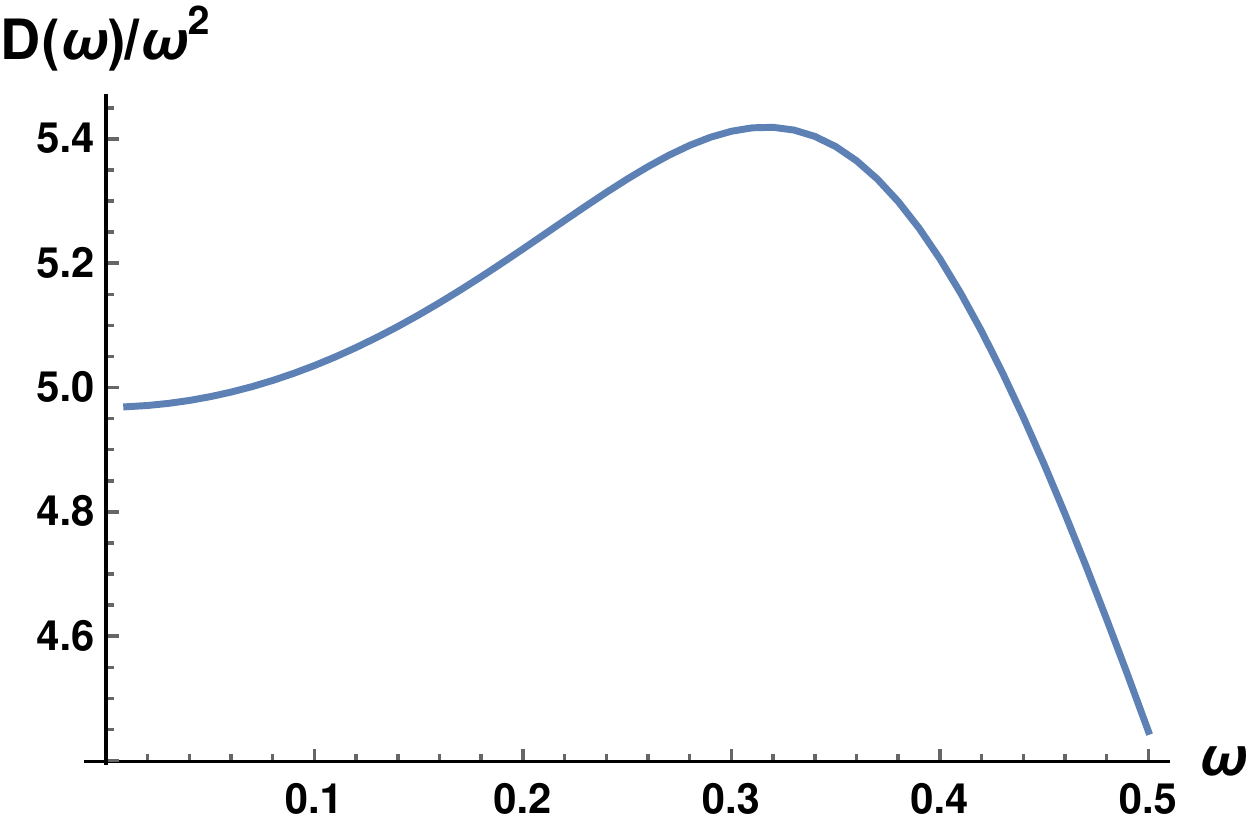}
\caption{Typical DOS in 3D, based on solving Eq. \eqref{eq:appendselfcon}. Parameters $k_D, d, c_0$ are chosen to be $1,0.1,0.5$ respectively. The height of curve has been rescaled.}
\label{fig:schirmacher}
\end{figure}

Figure \ref{fig:schirmacher} shows a typical plot of DOS calculated in this way, where $k_D=q_D=1$ and $c_0=0.5$. For convenience, we let the prefactor $(3N\gamma/2k_D^3)$ on the RHS in Eq. \eqref{eq:appendselfcon} be $d$.

\section{Dynamical structure factor}
According to \cite{Schirmacher2015, Cui2019}, the 3D longitudinal dynamical structure factor $S_L(k,\omega)$ has the following expression
\begin{equation}
S_L(k,\omega)=\frac{1}{\pi}[n(\omega)+1]\frac{k^2}{2\omega}\frac{k^2\text{Im}\Sigma(\omega)/\omega}{[\frac{k^2c_L^2(\omega)}{2\omega}-\frac{\omega}{2}]^2+[k^2\text{Im}\Sigma(\omega)/\omega]^2}
\end{equation}
where $n(\omega)+1=[1-\exp(-\hbar\omega/k_BT)]^{-1}$ is the Bose factor, $c_L(\omega)$ is the (generalized) longitudinal sound speed. 
In the classical limit, $\hbar\omega/k_BT\rightarrow0$, and using the fact that Im$\Sigma(\omega)\sim-\omega^2\ln(\omega b)$, we have
\begin{equation}
S_L(k,\omega)\propto\frac{k^2}{\omega^2}\frac{k^2(-\omega^2\ln(\omega b))/\omega}{[\frac{k^2c_L^2(\omega)}{2\omega}-\frac{\omega}{2}]^2+[k^2\omega^2\ln(\omega b)/\omega]^2}.
\label{append:dsf}
\end{equation}

\begin{figure}[h]
\centering
\includegraphics[width=7.5cm]{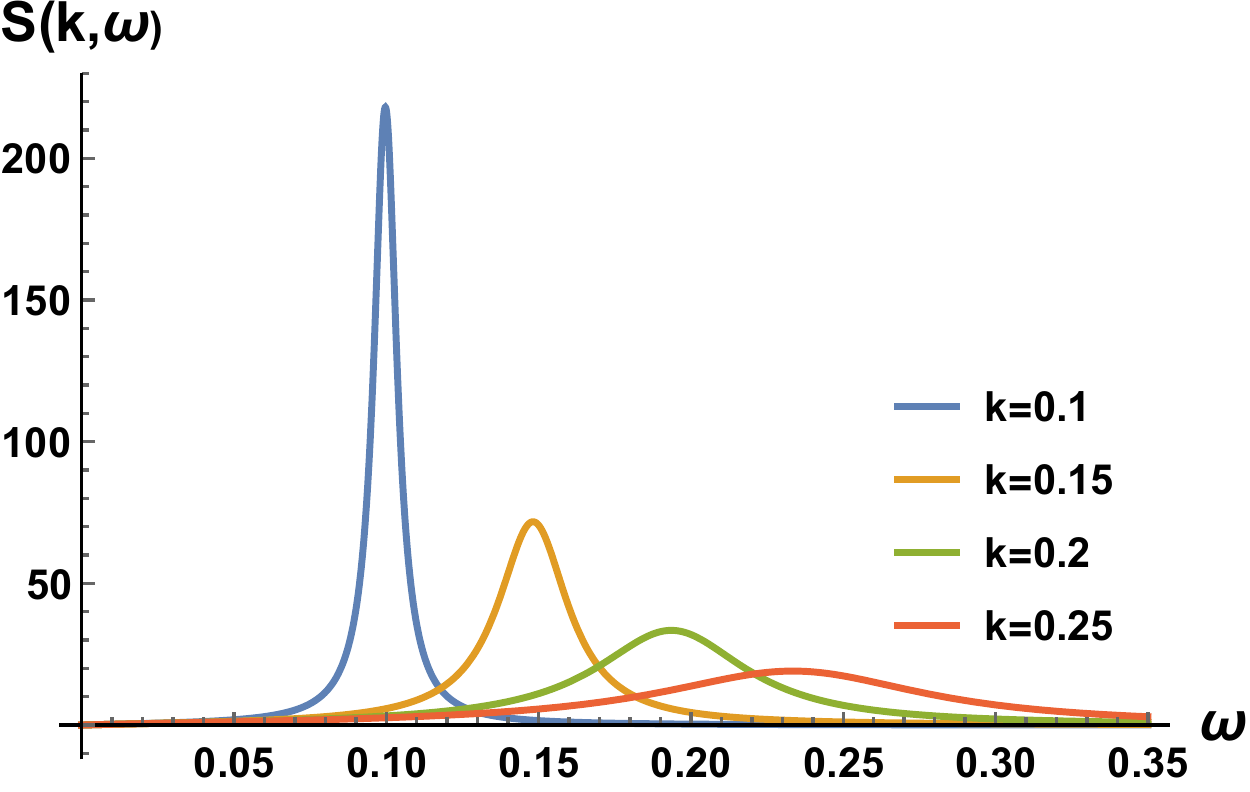}
\caption{Longitudinal dynamical structure factor $S(k,\omega)$ from Eq. \eqref{append:dsf}. The peak positions identifies resonance frequencies for the longitudinal acoustic excitations. The acoustic frequencies are found to correlate linearly with the wavevector $k$. Other parameters are the same as in Fig. \ref{fig:self3Dg} in the main text, namely $b=0.1$ and $c_L=1$.}
\label{fig:dsf}
\end{figure}

Taking same parameters as in Fig. \ref{fig:self3Dg} in the maintext, we show the longitudinal dynamical structure factor in Fig. \ref{fig:dsf}, where the linear dispersion relation is evident between the peak position and wavevector $k$. 

At low $\omega$, Eq. \eqref{append:dsf} can be fitted with a damped harmonic oscillator (DHO) model:
\begin{equation}
S(k,\omega)\propto\frac{k^2}{\omega^2}\frac{\Omega(k)^2\Gamma(k)}{(\omega^2-\Omega(k)^2)^2+\omega^2\Gamma(k)^2},
\end{equation}
where $\Omega(k)$ corresponds to the excitation frequency and $\Gamma(k)$ is the width of the Brillouin line (full width at half-maximum of the excitations). This is consistent with the proportionality coefficient between peak position frequency and $k$ identifying the longitudinal speed of sound $(c_L=1)$.
\end{appendix}

\bibliographystyle{apsrev4-1}

\bibliography{DOS}








\end{document}